\documentclass[journal=jacsat,manuscript=article]{achemso}
\setkeys{acs}{articletitle = true}

\usepackage[version=3]{mhchem} 
\usepackage{amsmath}  
\usepackage{amsfonts} 
\usepackage{subcaption}
\usepackage{graphicx} 
\usepackage{booktabs}
\usepackage{threeparttable}
\usepackage[font=footnotesize]{caption}
\usepackage{verbatim}
\usepackage{xcolor}
\usepackage{soul}
\usepackage{comment}
\usepackage{xr}
\usepackage{pdflscape}
\usepackage{tikz}
\usepackage{geometry}
\usepackage{dsfont}
\usetikzlibrary{matrix,positioning,decorations.pathreplacing} 
\SectionNumbersOn
\mciteErrorOnUnknownfalse
\makeatletter
\newcommand*{\addFileDependency}[1]{
  \typeout{(#1)}
  \@addtofilelist{#1}
  \IfFileExists{#1}{}{\typeout{No file #1.}}
}
\makeatother

\newcommand*{\myexternaldocument}[1]{%
    \externaldocument{#1}%
    \addFileDependency{#1.tex}%
    \addFileDependency{#1.aux}%
}

\myexternaldocument{supplement}

\title{A Localized-Orbital Energy Evaluation for Auxiliary-Field Quantum Monte Carlo}

\author{John L. Weber}
\email{jlw2245@columbia.edu}
\author{Hung Vuong}
\author{Pierre A. Devlaminck}
\affiliation{Department of Chemistry, Columbia University, 3000 Broadway, New York, NY, 10027}
\author{James Shee}
\affiliation{Kenneth S. Pitzer Center for Theoretical Chemistry, Department of Chemistry, University of California,
Berkeley, California 94720, USA}
\author{Joonho Lee}
\author{David R. Reichman}
\author{Richard A. Friesner}
\email{raf8@columbia.edu}
\affiliation{Department of Chemistry, Columbia University, 3000 Broadway, New York, NY, 10027}
\date{Sep 2021}

\begin{document}

\maketitle
\begin{abstract}
Phaseless Auxiliary-Field Quantum Monte Carlo (ph-AFQMC) has recently emerged as a promising method for the production of benchmark-level simulations of medium to large-sized molecules, due to its accuracy and favorable polynomial scaling with system size. Unfortunately the memory footprint of standard energy evaluation algorithms are non-trivial, which can significantly impact timings on graphical processing units (GPUs) where memory is limited. Previous attempts to reduce scaling by taking advantage of the low rank structure of the Coulombic integrals have been successful, but are significantly limited by high prefactors, rendering the utility limited to very large systems. Here, we present a complementary, cubic scaling route to reduce memory and computational scaling based on the low rank of the Coulombic interactions between localized orbitals, focusing on the application to phaseless AFQMC. We show that the error due to this approximation, which we term Localized Orbital AFQMC (LO-AFQMC), is systematic and controllable via a single variable, and is computationally favorable even for small systems. We present results demonstrating a robust retention of accuracy versus both experiment and full ph-AFQMC for a variety of test cases chosen for their potential difficulty for localized orbital based methods, including the singlet-triplet gaps of polyacenes benzene through pentacene, the heats of formation for a set of platonic hydrocarbon cages, and the total energy of ferrocene (Fe(Cp)$_2$). Finally, we reproduce our previous result of the gas phase ionization energy of Ni(Cp)$_2$, agreeing with full ph-AFQMC to within statistical error while using less than a fifteenth of the computer time.
\end{abstract}

\section{Introduction}

Solving the electronic structure of large, strongly correlated molecules at zero temperature represents a quintessential challenge of quantum chemistry. Quantum Monte Carlo methods, in particular projector-based Monte Carlo techniques such as Auxiliary-Field Quantum Monte Carlo (AFQMC), have emerged as promising approaches capable of producing benchmark results for small, strongly correlated systems.\cite{williams2020direct} This method relies on solving the Schrodinger equation in imaginary time $\tau$, projecting out the exact ground state $|\phi_0 \rangle$ under the requirement that the initial state $|\phi_i \rangle$ has a non-zero overlap, $\langle \phi_i | \phi_0 \rangle \neq 0$

\begin{equation}
    |\phi_0 \rangle \propto \lim_{\tau\to\infty} e^{-\tau \hat{H}} | \phi_i \rangle.
\end{equation}
AFQMC does so by reformulating the many body propagator via a Hubbard-Stratonovich transformation to an integral of a set of one body propagators coupled to auxiliary fields. This large multi-dimensional integral is estimated by Monte Carlo sampling a random walk of non-orthogonal Slater determinants, called walkers. Each walker can be propagated independently, aside from occasional communication to maintain efficient sampling populations (see later discussion on population control), and so this algorithm is nearly embarrassingly parallel. This allows one to make effective use of graphical processing units (GPUs), in favorable cases resulting in a $\simeq200\times$ reduction in the prefactor. Aside from a few model systems, however, one encounters a fermionic sign problem (or phase problem in the case of AFQMC). This leads to an exponential decrease in the signal to noise ratio, and thus an exponential increase in the required computational effort to obtain statistically meaningful results as the system grows. Algorithms for which the phase problem is uncontrolled, i.e. free projection AFQMC (fp-AFQMC), are formally exact, and the use of systematically better trial wavefunctions can reduce variance and allow convergence of exact results for small systems.\cite{mahajan2021taming} In order to render AFQMC polynomially scaling, the phase problem can be avoided via the phaseless approximation (ph-AFQMC), where we remove any phase accumulated in the weight by imposing a boundary condition based on the trial wavefunction.\cite{zhang2005quantum} This removes the phase problem and allows extended simulations of realistic materials, at the expense of a systematic bias with respect to the trial wavefunction. 

 The success of ph-AFQMC for \textit{ab initio} systems is in large part dependent upon the ability to systematically increase the quality of the results by increasing the quality of the trial wavefunction. Thus a fundamental question in the field centers on the optimal construction of trial functions, and there have been many studies exploring the space of quantum chemistry methods for efficient and accurate trial selection. In some cases, mean-field trial wavefunctions such as Hartree-Fock (HF) or Kohn-Sham Density Functional Theory (KS-DFT) are sufficient.\cite{shee2019singlet,weber2021silico,lee2020utilizing} In many cases, however, multi-determinant trials are necessary, particularly in transition metal containing complexes, with Configuration Interaction (CI) expansions including Complete Active Space Self-Consistent Field (CASSCF),\cite{weber2021silico,lee2020performance,shee2019achieving,rudshteyn2020predicting,rudshteyn2021calculation,purwanto2016auxiliary,purwanto2015auxiliary,purwanto2009excited,landinez2019non} Non-Orthogonal CI (NOCI),\cite{landinez2019non} and Semi-stochastic Heat-bath CI (SHCI) having been used for this purpose.\cite{mahajan2021taming,mahajan2022selected} Interestingly, even in some transition metal systems which are not expected to exhibit strong static correlation effects, multi-determinant trials have been necessary for ph-AFQMC to converge to within experimental results.\cite{rudshteyn2021calculation} However, use of a multideterminantal trial function leads to increased costs in both generating the trial and using it in AFQMC propagation and energy evaluation.  Addressing these issues is a key challenge inherent in the development of a scalable AFQMC approach with benchmark accuracy across a wide range of chemical problems.

For transition metal containing molecules, we have found that it is often necessary to use a multi-determinantal trial such as one provided by a CASSCF wavefunction (AFQMC/CAS) with a limited number of determinants ($N_{det} \simeq 300-1000$), in order to converge the result to chemical accuracy or near chemical accuracy.\cite{rudshteyn2020predicting,shee2019achieving,rudshteyn2021calculation} However, the use of AFQMC/CAS has been limited to small, single transition-metal systems ($<$ 1000 basis functions, 200 electrons), in part due to the enhanced computational cost and scaling of the local energy evaluation. Optimization of this step, therefore, may allow for the extension of AFQMC to larger transition metal systems of importance in biology, catalysis, and materials science.

For single determinant trials, the propagation of a walker within ph-AFQMC algorithms exhibits cubic scaling, whereas the energy evaluation exhibits quartic scaling. The most expensive steps in the propagation are the formation of the propagator, as well as the formation of the a shift in the auxiliary fields necessary for efficient importance sampling,\cite{zhang2003quantum} both of which scale as $XM^2$, or asymptotically $\mathcal{O}(M^3)$, where $M$ is the full basis dimension and $X$ is defined as the number of auxiliary fields, $X \simeq 4-10M$ for Cholesky decomposition thresholds between $10^{-4}$ and $10^{-5}$ Ha.\cite{motta2017computation} In the most straightforward implementation,\cite{shi2021some} the energy evaluation has a cost of $\mathcal{O}(N^2M^2)$, where N is the number of electrons in the molecule. There have been numerous attempts to decrease the scaling of the energy evaluation for single determinant AFQMC. Many take advantage of the low-rank nature of the ERIs beyond that of the literature standard, namely the modified Cholesky decomposition. For example, Motta \textit{et al.} formed a low-rank compression of each individual Cholesky matrix for single determinant trials.\cite{motta2019efficient} Additionally, Tensor Hyper-Contraction (THC) has emerged as a cubically scaling method with only quadratically scaling memory usage, albeit with a large prefactor that limits the utility of the approach to systems with $M > 2000$ to 4000.\cite{malone2018overcoming,malone2020accelerating} In addition to these low rank algorithms, Lee and Reichman applied a stochastic resolution of the identity (sRI) to reduce the scaling of the sum over Cholesky matrices, resulting in a cubic scaling algorithm with minimal overhead, although the accuracy has not been demonstrated for correlated systems.\cite{lee2020stochastic} It is also possible to take advantage of the sparsity of the Hamiltonian, where in a specified representation only elements above a certain threshold are explicitly stored. However, linear operations on sparse matrices are fairly inefficient, particularly on graphical processing units (GPUs), and thus doing so is only favorable in cases with extreme sparsity where nearly 90$\%$ of the elements are removed, which is in our experience typically not the case for molecular systems. 

When extending the trial to include multiple determinants, there are many choices of algorithms which take advantage of the excitation structure of CI expansions. The simplest involves using the Sherman Morrison Woodbury (SMW) algorithm to update the overlap between the walker and trial determinants.\cite{shee2018gpu} This approach significantly reduces the computational cost of multideterminantal propagation, but results in a scaling of $\mathcal{O}(N_{det}N^2M^2)$ for the energy evaluation, where $N_{det}$ is the number of determinants in the trial, resulting in a quickly intractable code for large determinant expansions. Recently, Mahajan \textit{et al.} have described an algorithm for the local energy evaluation of AFQMC with multi-determinant trials, which effectively reduces the scaling dependence on the number of determinants to $\mathcal{O}(NM^3 + N_{det}M)$.\cite{mahajan2021taming,mahajan2020efficient,mahajan2022selected} This algorithm enables the use of extremely accurate trial wavefunctions based on large SHCI expansions, which in turn reduces the variance of free-projection AFQMC enough to treat significantly larger systems than previously possible. While this opens the door to the usage of larger determinant expansions, for large systems the energy evaluation is still quartic, and thus remains the effective bottleneck. It is possible to modify THC to use within this framework, but this is again subject to the associated large prefactor and is likely to be useful only for very large systems.

Here we report an alternative (and in some cases complementary) method, directly applicable to both single determinant and multideterminant CI expansion trials, which takes advantage of the low rank of the electron repulsion integrals (ERIs) when working in a localized orbital basis. It does so by compressing the block of the ``half-rotated" ERI tensor corresponding to the interaction between each electron pair $[ij]$ using singular value decomposition (SVD), which is systematically controllable by an energetic threshold in addition to being compatible with dense linear algebra routines. This results in a reduced memory scaling from quartic to cubic for this tensor, with a concomitant reduction of scaling for the energy evaluation for a single determinant. We implement and outline the use of these localized ERIs within the SMW algorithm for multideterminant CAS trials, resulting in an energy evaluation algorithm that scales as $\langle M_{SVD}\rangle(N^2M + N_{det}N^2)$, where $\langle M_{SVD}\rangle$ represents the average rank of the compressed [ij] block, and we omit other values which are constant with system size, such as the maximum number of excitations from the reference for the set of determinants, $\epsilon$. This low rank localized structure is additionally compatible with the modified generalized Wick's theorem approach of Mahajan \textit{et al.}, and we outline one such algorithm in the SI, which results in a theoretical scaling of $\langle M_{SVD}\rangle N^2MA + N_{det}$, where A is the size of the active space. We move on to show the accuracy of Localized Orbital AFQMC (LO-AFQMC) in population control ph-AFQMC to within the statistical error of ph-AFQMC ($\simeq<$1 mHa) for a variety of representative molecules of nontrivial size and complexity, including polyacenes, metallocenes, and a set of benchmark platonic hydrocarbon cages.

We emphasize that although we focus on the application of this localized orbital compression to a particular set of algorithms within ph-AFQMC, much of this work is applicable to other orbital based Quantum Monte Carlo methods. We expect that the relative simplicity of the approach described here will enable these extensions. Due to the many advantages of GPU parallelization, we make efforts in this work to tailor our algorithm towards GPU architectures by maintaining an emphasis on both the reduction of memory requirements and the ability to perform operations efficiently in parallel using standard CUDA libraries.\cite{cuda}

The manuscript is organized as follows: In Sec. 2 we provide a brief overview of AFQMC (for a more in depth discussion see Ref \citenum{motta2017computation}), along with relevant multi-determinant algorithms for the local energy evaluation. In Sec. 3 we present the localized-orbital based algorithm, including some discussion of optimizing an implementation on GPU architectures. In Sec. 4 we then present the accuracy, memory usage, and timings for a series of test molecules. Lastly, we discuss the impact of such results, and conclude with our current outlook for AFQMC calculations in medium to large systems with strong correlation.

\section{Theory}
\subsection{Overview of ph-AFQMC}
Here we give a brief introduction to the framework of AFQMC; for a more in depth review of the method, we suggest some recent reviews.\cite{shi2021some,motta2018ab} In AFQMC, initial states are propagated in imaginary time $\tau$ according to
\begin{equation}
|\phi(\tau + \Delta\tau) \rangle = e^{-\Delta\tau \hat{H}} | \phi(\tau) \rangle = \int d\mathbf{x} P(\mathbf{x}) \hat{B}(\mathbf{x}) |\phi(\tau)\rangle \approx \sum_w \hat{B}(\mathbf{x}_w)|\phi_w(\tau)\rangle,
\end{equation}

where $\hat{H}$ denotes the electronic Hamiltonian, $\hat{H} = \hat{H}_1 + \hat{H}_2 = \sum_{pq} h_{pq} c^{\dag}_p c_q + \frac{1}{2} \sum_{pqrs} V_{pqrs} c^{\dag}_p c^{\dag}_q c_s c_r$, $|\phi_w\rangle$ denotes the Slater determinant associated with walker $w$, and $\hat{B} | \phi_w (\tau) \rangle = | \phi_w (\tau  + \Delta\tau) \rangle$ is another Slater determinant with rotated orbitals as given by the Thouless theorem.\cite{thouless1960stability} $V_{pqrs}$ are the two electron integrals referred to as $(pr|qs)$ and $\langle pq|rs \rangle$ in chemists and physicists notation, respectively.  For long imaginary times, computing observables using this Monte Carlo representation of the wavefunction will recover ground-state properties as long as the initial wavefunction has a non-zero overlap with the ground state. Using a symmetric Suzuki-Trotter decomposition, we separate the one- and two-body terms in $\hat{H}$ with an error quadratic in the imaginary time $\tau$,
\begin{equation}
    e^{-\Delta\tau(\hat{H}_1 + \hat{H}_2)} \simeq e^{-\frac{\Delta\tau\hat{H}_1}{2}}e^{-\frac{\Delta\tau\hat{H}_2}{2}} e^{-\frac{\Delta\tau\hat{H}_1}{2}} + \mathcal{O}(\Delta\tau^3).
\end{equation}
In practice, we mitigate the Trotter error by restricting our calculations to a small timestep, $\Delta\tau$ (0.005 Ha$^{-1}$ in this work). If we write the electronic two-body operator as a sum of one-body operators squared, $V_{pqrs} = \sum_{\alpha} L_{pr,\alpha}L_{qs,\alpha}$, which can be accomplished exactly via diagonalization or approximately via a density fitting or a modified Cholesky decomposition,\cite{purwanto2011assessing} we can then use the Hubbard-Stratonovich identity to convert the two-body operators into a multi-dimensional integral over a set of fluctuating ``auxiliary-fields'' x$_{\alpha}$,
\begin{equation}
    e^{-\frac{\Delta\tau}{2}(\sum_{\alpha} L_{\alpha}^2)} = \prod_{\alpha}{\int_{-\infty}^{\infty} \frac{1}{\sqrt{2\pi}} e^{-\frac{x_{\alpha}^2}{2}} e^{\sqrt{\Delta\tau}x_{\alpha}L_{\alpha}}dx_{\alpha}} + \mathcal{O}(\Delta\tau^2).
    \label{hubstrat}
\end{equation}
It is this multi-dimensional integral on which we perform Monte Carlo sampling
\begin{equation}
 |\phi(\tau + \Delta\tau) \rangle=\prod_{\alpha}\int_{-\infty}^{\infty} \frac{1}{2\pi} e^{-\frac{x_{\alpha}^2}{2}} e^{\sqrt{\Delta\tau} x_{\alpha}L_{\alpha}}dx_{\alpha}|\phi(\tau)\rangle= \int d\mathbf{x} P(\mathbf{x}) \hat{B}(\mathbf{x}) |\phi(\tau)\rangle,
 \label{endMC}
\end{equation}
where $\mathbf{x}$ is the vector of auxiliary fields. This Monte Carlo simulation can be reformulated as a branching, open-ended ensemble of random walkers $w$ over the manifold of Slater determinants, each represented by a single Slater determinant $\phi_{\tau,w}$ and corresponding weight $W_{\tau,w}$ and overlap with the trial $\langle \Phi_T |\Phi_{\tau,w} \rangle$. As each walker is propagated forward by $\hat{B}(\mathbf{x}_{\tau,w})$, with the space of auxiliary fields \textbf{x} being sampled from the Gaussian probability defined in Eq. \ref{hubstrat}, the weights are updated according to the ratio of the new overlap with the trial to the old overlap
\begin{equation}
    |\Phi_{\tau + \Delta\tau,w} \rangle = \hat{B}(\mathbf{x}_{\tau,w}) |\Phi_{\tau,w} \rangle,
\end{equation}
\begin{equation}
    W_{\tau + \Delta\tau,w} e^{i\theta_{\tau + \Delta\tau,w}} = \frac{\langle \Phi_T  | \Phi_{\tau + \Delta\tau,w} \rangle}{\langle \Phi_T |\Phi_{\tau,w} \rangle}W_{\tau,w} e^{i\theta_{\tau,w}}.
\end{equation}

All theory thus far is formally exact; however, the fermionic phase problem leads to an exponential decrease in signal-to-noise ratio as the walkers are propagated, resulting in an exponentially growing population of walkers (and thus computational time) necessary to achieve a given statistical error. In order to mitigate this noise, we perform importance sampling by shifting the auxiliary fields to favor sampling in regions with high(er) overlap with the trial wavefunction.\cite{zhang2003quantum,rom1998shifted} This is complemented by the phaseless constraint, in which we multiply each walker's weight by a factor corresponding to projecting the accumulated phase with respect to the trial wavefunction back onto the real axis, namely we multiply by max(0, cos($\Delta\theta$)). 

Within both free projection and phaseless AFQMC, there are numerous ways to perform the sampling over auxiliary fields. It is advantageous for computational efficiency to restrict the number of walkers to some set value. This is most typically done by periodically annihilating walkers with low weights and duplicating those with high weights with a probability proportional to the distance of the walker's weight from one, and keeping the total number of walkers constant, referred to as Population Control (PC). PC introduces a bias which scales linearly with the inverse number of walkers, which is typically minimal when running with hundreds to a few thousands of walkers, and in the absence of the phase problem leads to stable calculations over hundreds of Ha$^{-1}$. In this work we implement population control via a ``comb'' algorithm,\cite{booth2009monte,buonaura1998numerical} doing so every 20 time steps, along with measurements of energy. Additionally, the walkers themselves need to be periodically orthonormalized for numerical stability; we do so every two steps. A mean field subtraction is performed prior to propagation to reduce variance.\cite{motta2017computation}

As an alternative to PC, we have recently introduced a sampling approach based on direct calculation of energy differences between two relevant states (e.g. +2 and +3 states to compute ionization potentials) which is designated “correlated sampling” (CS).\cite{shee2017chemical} 
For problems which can be effectively formulated to utilize CS, i.e. similar geometries, it can provide an attractive combination of speed and accuracy, at least for the systems tested to date.\cite{rudshteyn2020predicting,rudshteyn2021calculation,shee2019achieving}
While we do not report values for free projection or correlated sampling in this work, we note that as the local energy evaluation remains identical for both, we expect LO-AFQMC to exhibit similar results.

\subsubsection{Evaluation of the Local Energy}
The local energy associated with a walker is defined as 
\begin{equation}
E = \frac{\langle \phi_T | \hat{H} | \phi \rangle}{\langle \phi_T | \phi \rangle}.
\end{equation}
The two-body contribution to the energy can be written as
\begin{equation}
E_2 = \frac{1}{2}\sum_{pqrs}^M  V_{pqrs} \frac{\langle \phi_T | a_p^{\dag} a_q^{\dag} a_s a_r |  \phi \rangle }{\langle \phi_T | \phi \rangle},
\end{equation}
which, when the generalized Wick's theorem is used, gives
\begin{equation}
E_2 = \frac{1}{2}\sum_{pqrs}^M  V_{pqrs} \sum_{\sigma \tau} (G_{p \sigma, r \sigma}G_{q \tau, s \tau} - G_{p \sigma, s \tau}G_{q \tau, r \sigma}),
\label{2e_general}
\end{equation}
where $\sigma$ and $\tau$ are spin indices. The equal-time Green's function is defined as 
\begin{equation}
G_{p \sigma, q \tau} = \frac{\langle \phi_T | a^\dag_{p \sigma} a_{q \tau} | \phi \rangle}{\langle \phi_T | \phi \rangle}  = \delta_{\sigma \tau} [\Phi ( \Phi_T^\dag \Phi )^{-1} \Phi_T^\dag]^{\operatorname{T}},
\end{equation}
where $\delta_{\sigma \tau}$ is the Kronecker delta function. Hereafter, indices $i,j$ run over the number of electrons $N$, and $p,q,r,s$ index the entire set of basis functions $M$.  $| \phi_T \rangle$, the trial wavefunction, constrains the paths of the random walk according to the phaseless constraint. $\Phi$ is used to denote the matrix representation of a Slater determinant, with columns representing orbitals, e.g. linear combinations of orthonormal basis functions.  These matrices have dimensions $M\times N$ ($M=$ basis size, $N=$ number of electrons). In the case of a CASSCF trial, $| \phi_T \rangle = \sum_d^{N_{det}} c_d | \phi^d_T \rangle$, where every $| \phi^d_T \rangle$ corresponds to a determinant with a maximum of $\epsilon$ excitations from the reference determinant. 

The expression for the two-body contribution to the local energy is given by
\begin{equation}
E_2 = \sum_d^{N_{det}}\frac{c_d\langle\Phi_T^d|\Phi\rangle}{ \sum_d^{N_{det}}c_d\langle\Phi_T^d|\Phi\rangle}
 \times \frac{1}{2}\sum_{pqrs}^M V_{pqrs} \sum_{\sigma \tau} (G^d_{p \sigma, r \sigma}G^d_{q \tau, s \tau} - G^d_{p \sigma, s \tau}G^d_{q \tau, r \sigma}),
\label{2e_RCAS}
\end{equation}
where
\begin{equation}
G^d_{p \sigma, q \tau} = \frac{\langle \phi^d_T | a^\dag_{p \sigma} a_{q \tau} | \phi \rangle}{\langle \phi_T^d | \phi \rangle}  = \delta_{\sigma \tau} [\Phi ( \Phi_T^{d\dag} \Phi )^{-1} \Phi_T^{d\dag}]_{qp}.
\label{G}
\end{equation}
In practice, the four index ERI tensor is represented as a three-index factorized tensor $V_{pqrs} = \sum_{\alpha}^{X} L_{pr}^{\alpha}L_{qs}^{\alpha}$, where $X$ ranges from 4$M$ to 10$M$ for Cholesky decompositions with thresholds between $10^{-4}$ and $10^{-6}$. A naive implementation of the local energy evaluation would thus scale as $\mathcal{O}(N_{det}M^4)$. In typical ph-AFQMC calculations the local energy must be evaluated approximately 4 to 6 million times; the energy evaluation can thus quickly become a computational bottleneck. However, there are many algorithmic tricks one can use that can significantly reduce the scaling of this step, allowing for the extension of AFQMC/CAS to treat large scale \textit{ab initio} systems in an accurate manner. We describe one such possibility for scaling reduction below.

\subsection{Half-Rotated ERIs}
We begin by recognizing that the trial wavefunction is known prior to propagation. We define an intermediate matrix, $Q$, which partially excludes the dependence of the Green's function on the trial
\begin{equation}
G^d_{p \sigma, q \tau} = 
\sum_i^N Q^d_{qi} \Phi_{T,ip}^{d\dag},
\label{GwQ}
\end{equation}
\begin{equation}
Q^d_{qi} = [\Phi ( \Phi_T^{d\dag} \Phi ) ^{-1}]_{qi} .
\label{Q}
\end{equation}
As both the trial wavefunction and ERIs are known, we can precompute the ``half-rotated" cholesky vectors, $\bar{L}_{ri}^{\alpha} =\sum_p^M L_{rp}^{\alpha} \Phi_{T,pi}$, at a cost of $XM^2N_{cas}$ at the beginning of the calculation, where $N_{cas}$ corresponds to the inactive occupied plus active orbitals (aka each orbital that can be occupied in a trial determinant). $N_{cas}$ can thus be as small as N (for a single determinant) or as large as M for a full CI expansion; for the case of interest here, relatively limited CASSCF active space CI expansions, $N_{cas} \simeq N$. Use of this precomputed tensor and the Q intermediates results in an algorithm which scales as $N_{det}XMN^2$
\begin{equation}
E_2 = \sum_d^{N_{det}}\frac{c_d\langle\Phi_d|\Phi\rangle}{ \sum_d^{N_{det}}c_d\langle\Phi_d|\Phi\rangle}\times
  \frac{1}{2}\sum_{ij}^{N_{cas}} \sum_{rs}^M \sum_{\alpha}^{X} \bar{L}_{ri}^{\alpha}\bar{L}_{sj}^{\alpha} \sum_{\sigma \tau} (Q^d_{r \sigma, i \sigma}Q^d_{s \tau, j \tau} - Q^d_{s \sigma, i \tau}Q^d_{r \tau, j \sigma}).
\label{2e_HRchol_RCAS}
\end{equation}

 The cost of this energy evaluation can further be reduced via precomputation of the sum over $X$, resulting in $\mathcal{O}(N_{det}M^2N^2)$ scaling at the cost of storing a $N_{cas}^2M^2$ dimensional tensor in memory, $Y_{ijrs} = \sum_{\alpha}^{X} \bar{L}_{ri}^{\alpha} \bar{L}_{sj}^{\alpha}$, giving
\begin{equation}
E_2 = \sum_d^{N_{det}}\frac{c_d\langle\Phi_d|\Phi\rangle}{ \sum_d^{N_{det}}c_d\langle\Phi_d|\Phi\rangle}\times
  \frac{1}{2}\sum_{ij}^{N_{cas}} \sum_{rs}^M Y_{ijrs} \sum_{\sigma \tau} (Q^d_{r \sigma, i \sigma}Q^d_{s \tau, j \tau} - Q^d_{s \sigma, i \tau}Q^d_{r \tau, j \sigma}).
\label{2e_HRERI_RCAS}
\end{equation}

Note that if $i$ or $j$ are in the active space of a CAS trial, they might not be present in every determinant. In order to effectively perform the sum over determinants in parallel, we find it advantageous to store the set of $Q^d$ matrices as if they include every orbital, substituting with zeros when this is not the case. While the HR-ERI algorithm is generally faster for small CI expansions and smaller systems in general, this approach can quickly present a bottleneck for large systems. This is especially apparent in calculations utilizing GPUs, due to the inability to store the half-rotated $Y_{ijrs}$ on a single GPU. This renders large (and expensive) interconnected GPU clusters a requirement, and results in the efficiency being primarily controlled by memory passing protocols, which can easily lead to major slowdowns for medium to large systems around M greater than around $1000$. When running on clusters with fast GPU to GPU transfer speeds, however, the effect can be minimal, and it is often then advantageous to precompute $Y_{ijrs}$ even in the case where it cannot fit on a single GPU.

\subsection{LO-AFQMC Algorithms}
We here present the general structure of localized ERIs, before outlining the application and use of localization to the HR-ERI algorithm.
\subsubsection{Compression of the HR-ERI Tensor}
We begin by noting that the structure of the half-rotated tensor, $Y_{ijrs}$, corresponds to $N_{cas}^2$ blocks of dimension $M \times M$ corresponding to the interaction integrals between the occupied orbital pair $[ij]$, which we denote $\{Y_{rs}^{[ij]}\}$. It is well known that the low-rank structure of the Coulombic integrals is best revealed when dealing with interactions between distinct localized orbitals.\cite{neese2009efficient} This low rank structure can be taken advantage of in numerous ways; we do so via a block-wise singular value decomposition for each localized orbital pair $[ij]$, which is then truncated according to a threshold beyond which singular values are discarded, here denoted T$_{SVD}$. The result of this procedure is an $N_{cas} \times N_{cas}$ list of rectangular matrices with dimensions $M \times M_{SVD}$ and $M_{SVD} \times M$ with $M_{SVD}$ being the truncated dimension. Thus, the expression for $Y_{rs}^{[ij]}$ may be written
\begin{equation}
Y_{rs}^{[ij]} \simeq \sum_{K^{[ij]}}^{M_{SVD}^{[ij]}} 
U^{[ij]}_{rK} \Sigma_K^{[ij]} V^{[ij]}_{Ks}
= \sum_{K}^{M_{SVD}} 
\bar{U}^{[ij]}_{rK} \bar{V}^{[ij]}_{Ks},
\end{equation}
where $U^{[ij]}$ and $V^{[ij]}$ are the set of unitary left and right singular vectors, and $\Sigma^{[ij]}$ are their associated singular values. For notational clarity, we omit the indices [ij] from $M_{SVD}$ and K throughout the remainder of the paper. At $M_{SVD} = M$ (i.e. T$_{SVD}$ = 0), this expression becomes exact. This compression is performed once in the beginning of the simulation, at a cost of $\mathcal{O}(N^2M^3)$, which is easily parallelized over GPUs for a speedup of $\frac{1}{N_{GPU}}$, and represents a minimal ($\simeq1$ to $2\%$) addition to the time of the calculations outlined in this study. We choose to allow $M_{SVD}$ to vary between [ij] pairs, instead fixing $T_{SVD}$ to obtain a given accuracy. It is additionally possible to employ approximate versions of the singular value decomposition, which would reduce the scaling of this step significantly, although we did not explore this in this work. In practice, $\langle M_{SVD} \rangle$, the average compressed dimension, does not scale with the system size (and in fact decreases with a given $T_{SVD}$ as a function of the system size for a given class of molecules, as we demonstrate in section \ref{results}), and thus the memory scaling is effectively reduced from quartic to cubic in system size, namely $N_{cas}^2M$.

\subsubsection{LO-AFQMC in the HR-ERI Algorithm}

We next explicitly express the energy for a CASSCF trial using the HR-ERI algorithm within the compressed framework using
\begin{equation}
E_2 = \sum_d^{N_{det}}\frac{c_d\langle\Phi_d|\Phi\rangle}{ \sum_d^{N_{det}}c_d\langle\Phi_d|\Phi\rangle}
 \times \frac{1}{2}\sum_{ij}^{N_{cas}} \sum_K^{M_{SVD}} \sum_{rs}^M  \bar{U}^{[ij]}_{rK} \bar{V}^{[ij]}_{Ks}
 (4Q_{r,i}^dQ_{s, j}^d - 2Q_{s,i}^dQ_{r,j}^d).
\label{2e_HR_RCAS_LO}
\end{equation}
The complexity for a direct summation is now $\mathcal{O}(N_{det}N^2M\langle M_{SVD} \rangle)$, and for a single determinant trial this is straightforward and results in $\mathcal{O}(N^2M\langle M_{SVD}\rangle)$ scaling. Due to the introduction of a low rank index K, however, we can extend the savings for the half-rotated algorithm further with respect to the number of determinants by taking advantage of the structure of the CAS trial. In the HR-ERI algorithm we use a Sherman-Morrison-Woodbury (SMW) algorithm to update the overlap of each walker with the determinants of the trial. In other words, we can write the $N_{det}$ intermediate Green's functions $Q_{ri}^d$ of Eq. \ref{Q} as a single contribution from the reference, along with rank $\epsilon$ corrections specific to each determinant
\begin{equation}
    Q_{ri}^d = \Phi*(A + U^d V^{d\operatorname{T}})^{-1} = 
    \Phi A^{-1} - \Phi A^{-1}U^d(I + V^{d\operatorname{T}} A^{-1} U^d)^{-1}V^{d\operatorname{T}} A^{-1},
    \label{Q_SMW_reg}
\end{equation}
where $A = \Phi_T^{0\dag}\Phi$, $\Phi_T^{0\dag}$ is the reference determinant, and $U^d$ and $V^{d,\operatorname{T}}$ are $N$ by $\epsilon$ matrices corresponding to the rows which are changed via excitation from the reference for a given determinant. From here on $\epsilon$ will refer to the maximum number of excitations possible in the CAS trial, which is typically $\simeq 6$ and bounded by the size of the active space. The scaling of the formation of the set of all $Q$ matrices has now been transformed from $\mathcal{O}(N_{det}N^2M)$ to $\mathcal{O}( N^2M + N_{det}N\epsilon^2)$ when in the molecular orbital basis (see Sec. \ref{SMW_GF_SI}). If one formed the full set of $N_{det}$ $N \times M$ matrices $Q_{ri}^d$, as in the full HR-ERI algorithm, the resulting operations in Eq. \ref{2e_HR_RCAS_LO}, computed in a loop over determinants, would scale as  $\mathcal{O}( N_{det}N^2M \langle M_{SVD} \rangle )$. However, If we rearrange the sums in Eq. \ref{2e_HR_RCAS_LO} so as to avoid the explicit computation of $Q_{ri}^d$, we can effectively replace another factor of the full basis dimension, $M$, with $\langle M_{SVD} \rangle$.

Since the LO-ERI are specific to the pairs $[ij]$, we must operate on a column of the $Q$ matrix, $Q_{r,i}^d$, corresponding to the specific orbital $i$,
\begin{equation}
    Q_{r,i}^d = \Phi A^{-1}_i - \Phi A^{-1} U^d (I + V^{d\operatorname{T}} A^{-1} U^d)^{-1} V^{d\operatorname{T}} A^{-1}_i.
    \label{Q_SMW}
\end{equation}
Note that the $U^d$ matrix has a block structure of zero rows followed by an identity (or permutation matrix) of maximum dimensions of $\epsilon \times \epsilon$, and so $A^{-1} U^d$ is equal to the last $\epsilon$ columns (or permutations thereof) of $A^{-1}$ and requires no computation. Additionally, note that each $Q$ has $N$ total columns, whereas the index $i$ in the energy evaluation (see Eq. \ref{2e_HR_RCAS}), is summed over $N_{cas}$; thus the index ``$i$'' in Eq. \ref{2e_HR_RCAS} corresponds to the index of the column of the $d$-th determinant in the original CAS MOs. This means that for a specific determinant, the $i$-th column might not exist - thus, to allow for efficient operations in parallel on GPUs (``batching"), which is most effective for processes with equivalent operation count, we assume that they do in memory, substituting with zeros if needed. 

We can now combine Eqns. \ref{2e_HRERI_RCAS} and \ref{Q_SMW}. This involves effectively performing the sum over full basis dimensions $r$ and $s$ first. We then form four vectors of dimension $\langle M_{SVD} \rangle$, $\bar{Q}_{K,L,i}^d = \bar{U}^{[ij],\operatorname{T}}_{rK}*Q_{ri}^d$, $\bar{Q}_{K,R,i}^d = \bar{V}^{[ij]}_{Ks}*Q_{si}^d$, $\bar{Q}_{K,L,j}^d = \bar{U}^{[ij],\operatorname{T}}_{rK}*Q_{rj}^d$, and $\bar{Q}_{K,R,j}^d = \bar{V}^{[ij]}_{Ks}*Q_{sj}^d$, for every $[ij]$ pair and every determinant. If we multiply by the determinant coefficients and overlaps in this step, the resulting sums over $[ij]$, determinants, and $K$, are simply a dot product (e.g. for the Coulomb interaction $\bar{Q}_{K,L,i}^{d,[ij]} * \bar{Q}_{K,R,j}^{d,[ij]}$, scaling as $\mathcal{O}(N_{det}N^2\langle M_{SVD} \rangle)$). Explicitly, we write
\begin{equation}
E_2 = \sum_d^{N_{det}}\frac{c_d\langle\Phi_d|\Phi\rangle}{ \sum_d^{N_{det}}c_d\langle\Phi_d|\Phi\rangle}
 \times \frac{1}{2} \sum_{ij}^{N_{cas}} \sum_K^{M_{SVD}}
 (4\bar{Q}_{K,L,i}^{d,[ij]}\bar{Q}_{K,R,j}^{d,[ij]}- 2\bar{Q}_{K,L,j}^{d,[ij]}\bar{Q}_{K,R,i}^{d,[ij]}),
\label{2e_HR_RCAS}
\end{equation}
\begin{equation}
    \bar{Q}_{K,L,i}^{d,[ij]} = \bar{U}^{[ij],\operatorname{T}}_{rK}*Q_{ri}^d = 
    \bar{U}^{[ij]}_{K,r}\Phi A^{-1}_i - \bar{U}^{[ij]}_{Kr}\Phi A^{-1} U^d(I + V^{d,\operatorname{T}} A^{-1} U^d)^{-1}V^{d,\operatorname{T}} A^{-1}_i
    \label{BigQ}
\end{equation}
Consider now the formation of $\bar{Q}_{K,L,i}^{d,[ij]}$, for which the scaling of key intermediates has been outlined in Table \ref{table:intermediates}. Note that for every $[ij]$ pair, we only require $\bar{U}^{[ij]}_{Kr}$ intersect with vectors $A^{-1}_i$ and $A^{-1}_j$, instead of the entire $M \times M$ matrix $A^{-1}$. For the reference section, i.e. the first term of the right hand side of Eq. \ref{BigQ}, the computational order is straightforward. Specifically, we store the full $\Phi A^{-1}$, then define vectors corresponding to $i$ and $j$, and lastly multiply $\bar{U}^{[ij]}_{Kr}$ by these vectors. This set of steps scales as $N^2 M \langle M_{SVD} \rangle$, and must occur four times, once for every $\bar{Q}$ in Eq. \ref{2e_HR_RCAS}. 

\begin{table}[!htb]
    \centering
    \begin{tabular}{c | c c c}
         Intermediate & Operation & Memory scaling & Computational scaling \\
         $\{Q^{\textrm{int}}_{L}\}_{[ij],d,K}$ & $\bar{U}^{[ij]}_{Kr} \times (\Phi A^{-1} U^d)$ &$N^2\langle M_{SVD} \rangle \epsilon$ & $N^2M\langle M_{SVD} \rangle \epsilon$\\
         $\{S\}_d$  &  $(I + V^{T,d} A^{-1} U^d)^{-1} \times V^{T,d}A^{-1}$ & $N_{det}N\epsilon$ & $N_{det}N\epsilon^2$ \\
         $\{\bar{Q}_{L,i}\}_{[ij],d,K}$ & $Q^{\textrm{int}}_{L} \times S_i$ & $N_{det} N^2 \langle M_{SVD} \rangle$ & $N_{det} N^2 \langle M_{SVD} \rangle \epsilon$
    \end{tabular}
    \caption{List of key intermediates in the LO-AFQMC evaluation of the 2-body energy using SMW, including the memory and computational scaling. The nomenclature $\{X\}_{a,b,c}$ indicates that the scaling listed is for the formation of the set of intermediates X for all valid indices a, b, c. }
    \label{table:intermediates}
\end{table}

For the SMW correction term (the rightmost term in Eq. \ref{BigQ}), we must take advantage of the block identity form of $U^d$, namely that the first $N_{inact}$ (occupied orbitals which are not in the active space) columns of $\Phi A^{-1}$ are zero regardless of the determinant. Thus we need only compute $Q^{\textrm{int}}_{L} = \bar{U}^{[ij]}_{Kr}\Phi A^{-1}_{act}$ once, which scales as $N^2 M \langle M_{SVD} \rangle \epsilon$,  where $A^{-1}_{act}$ is the last $\epsilon$ columns of $A^{-1}$. This must be performed twice, once for $\bar{U}^{[ij]}_{Kr}$ and once for $\bar{V}^{[ij]}_{Ks}$. 

If one generates the $N_{det}$ intermediates of size $\epsilon\times N$, corresponding to 
$\{S\}_d = (I+V^{d,\operatorname{T}}A^{-1}U^d)^{-1}V^{d,\operatorname{T}}A^{-1}$ (see details in Sec. \ref{SMW_GF_SI}), it is straightforward to determine the column of these intermediates corresponding to $i$ for each determinant and then multiply the stored $Q^{\textrm{int}}_{L}$ by these column vectors and subtract them from the reference value. This series of steps then scales as $N_{det} N^2 \langle M_{SVD} \rangle \epsilon$. The total scaling of the HR-ERI energy evaluation is thus reduced from $(N_{det}N^2M\langle M_{SVD} \rangle + N_{det}N^2\langle M_{SVD} \rangle)$ to $(N^2 M \langle M_{SVD} \rangle(\epsilon + 1) + N_{det} N^2 \langle M_{SVD} \rangle \epsilon)$. Keeping only dimensions that scale with the system size, this is thus an asymptotically $\mathcal{O}(N^2M + N_{det}N^2)$ scaling algorithm. While the use of SMW to speed up calculations has been well documented, we additionally provide details with respect to the generation of the Green's function in the MO basis, for 1-body energetic terms and the evaluation of the force bias in Sec. \ref{SMW_GF_SI}.

\section{Results}\label{results}
Here we present results illustrating the efficiency and accuracy of the LO-AFQMC approach. The specifics of the localization method used were not observed to significantly affect results, thus only a Foster-Boys localization scheme was used to produce the results shown here.\cite{foster1960canonical} Only the inactive orbitals were localized; localization of the active orbitals after CASSCF iterations reduced the efficiency of the CI expansion, resulting in orders of magnitude more determinants being required in the CI expansion to produce the same total sum of CI weight, and thus a loss of efficiency. 

\begin{figure}[H]
    \centering
    \includegraphics[width=10cm]{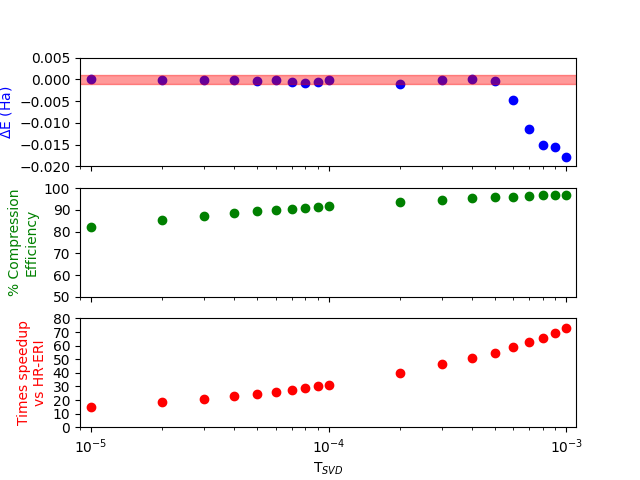} 
    \caption{Convergence of the LO-AFQMC error with regards to T$_{SVD}$ for dodecane in the cc-pVDZ basis. The error vs regular AFQMC (top, blue) converges to within 1 mHa (red shaded region) around T$_{SVD}$ = 0.0005. Note that values to the left have a tighter threshold, and that the x axis is in a logarithmic scale. The compression efficiency (middle, green) ranges between 82 and 97 $\%$, with an intermediate $T_{SVD} = 10^{-4}$ threshold reducing M by 91.75$\%$. The reduced computational scaling due to the compression is reflected in the speedup vs regular HR-ERI (bottom, red) of the energy evaluation for 16 walkers, which even for the tightest threshold tested (T$_{SVD}$ = 10$^{-5}$) is over 15x faster than our HR-ERI code, with $T_{SVD} = 10^{-4}$ being $\simeq 31$ times faster. These calculations were performed on a single NVIDIA GeForce RTX 3090.
    } 
    \label{figure:dodecane_convergence}
\end{figure}

We first investigate dodecane in a cc-pVDZ basis as a medium sized test case. In order to estimate the error due to compression of the HR-ERI tensor, we take the difference in energies between that arising from the full and localized HR-ERI tensors to evaluate the local energy of a single walker (Figure \ref{figure:dodecane_convergence}). The error estimated in this manner converges with respect to the threshold T$_{SVD}$, with T$_{SVD}$ = 0 reproducing the full AFQMC result. For ph-AFQMC, our goal is simply to reduce systematic errors to below the statistical error that one is seeking, which is generally around 1 mHa for molecular systems. This is achieved for a single energy evaluation for dodecane around T$_{SVD} = 5 \times 10^{-4}$, with an overall compression efficiency ($\frac{M - \langle M_{SVD} \rangle}{M} * 100$) of 96 percent. We note that the reduction in error with $T_{SVD}$ does not occur monotonically.
Error reduction does decrease systematically, however, and the magnitude of fluctuations at any given threshold are proportional to the threshold itself. To assess the increased scaling due to this compression, we performed calculations for 16 walkers on a single GPU, and present the speedup versus storing and using the full HR-ERI tensor. Even the tightest threshold tested (T$_{SVD}$ = 10$^{-5}$) is over 15x faster than our HR-ERI code for a single energy evaluation. 

\begin{figure}[H]
    \centering
    \includegraphics[width=18cm]{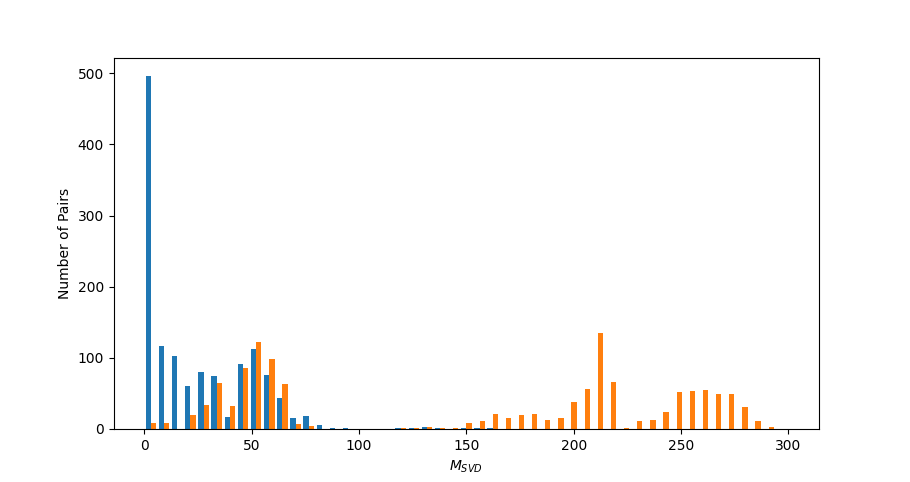} 
    \caption{A histogram of M$_{SVD}$ for LO-AFQMC in the canonical (orange) and localized (blue) molecular orbital bases for dodecane with T$_{SVD} = 10^{-4}$. Localization results in an increase in compression efficiency from 48.89$\%$ to 91.75$\%$. Note that the full basis size M = 298.
    } 
    \label{figure:dodecane}
\end{figure}

To illustrate the necessity of using localized orbitals, we compare LO-AFQMC (T$_{SVD}$ = $10^{-4}$) using both localized and non-localized orbitals for dodecane (Fig. \ref{figure:dodecane}). It can be seen that the degree of compression is significantly reduced when using the canonical basis, highlighting the intuitive fact that localized and distant orbital pairs exhibit low rank blocks within the HR-ERIs, similar to observations in other local correlation methods.\cite{neese2009accurate}

\subsection{Metallocenes}

\begin{figure}[H]
    \centering
    \begin{minipage}[c]{0.5\textwidth}
    \centering
    \includegraphics[width=0.3\textwidth]{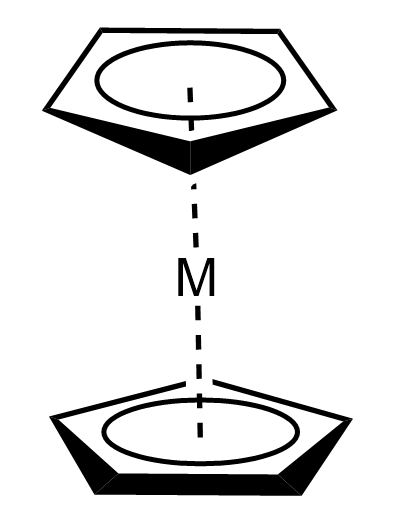}
  \end{minipage}\hfill
  \begin{minipage}[c]{0.5\textwidth}
    \caption{
       The chemical structure of a generic metallocene, M(Cp)$_2$
    } \label{figure:metallocene}
  \end{minipage}
\end{figure} 

While our results for dodecane are promising, it is necessary to test whether or not this localization scheme works for more challenging molecules where AFQMC may be a method of choice. Recently, we presented ph-AFQMC data on the gas phase ionization energy of a series of metallocenes (Fig. \ref{figure:metallocene}), for which we found it necessary to use multideterminant CASSCF trials. Here, we use calculations on ferrocene as benchmark results for LO-AFQMC, using geometries from Ref. \citenum{rudshteyn2021calculation}. Population control calculations for Fe(III)(Cp)$_2$ were run with a series of T$_{SVD}$ values and compare our results to the full AFQMC total electronic energies in a cc-pVTZ-dkh basis (Table \ref{table:ferrocene}). We again find T$_{SVD} = 10^{-4}$ to be a good choice to reduce the error in the total energy to the level of statistical error, while still maintaining a compression efficiency of 77.5$\%$. We hereafter then use T$_{SVD} = 10^{-4}$ as the default threshold for the remaining results, although one might need to increase the threshold if one desires an increased accuracy.

\begin{table}
\centering
\caption{Errors in total energy for ferrocene (cc-pVTZ-dkh) vs an identical AFQMC calculation with full ERIs for a series of SVD truncation thresholds T$_{SVD}$. T$_{SVD} = 10^{-4}$ provides a good balance between memory efficiency and accuracy.}
\begin{tabular}{l r r}
\hline
T$_{SVD}$	&	$\Delta$E(Ha)	&	$\%$ Compression	\\ 
\hline
0.00005	&	0.0003(13)	&	70.84	\\
0.0001	&	0.0010(10)	&	77.50	\\
0.00025	&	0.0013(15)	&	85.04	\\
0.0005	&	0.0034(11)	&	89.52	\\
0.00075	&	0.0049(11)	&	91.64	\\
0.001	&	0.0086(11)	&	92.91	\\
\hline
\end{tabular}
\label{table:ferrocene}
\end{table}

\subsection{Polyacenes}

\begin{figure}[H]
    \centering
    \includegraphics[width=12cm]{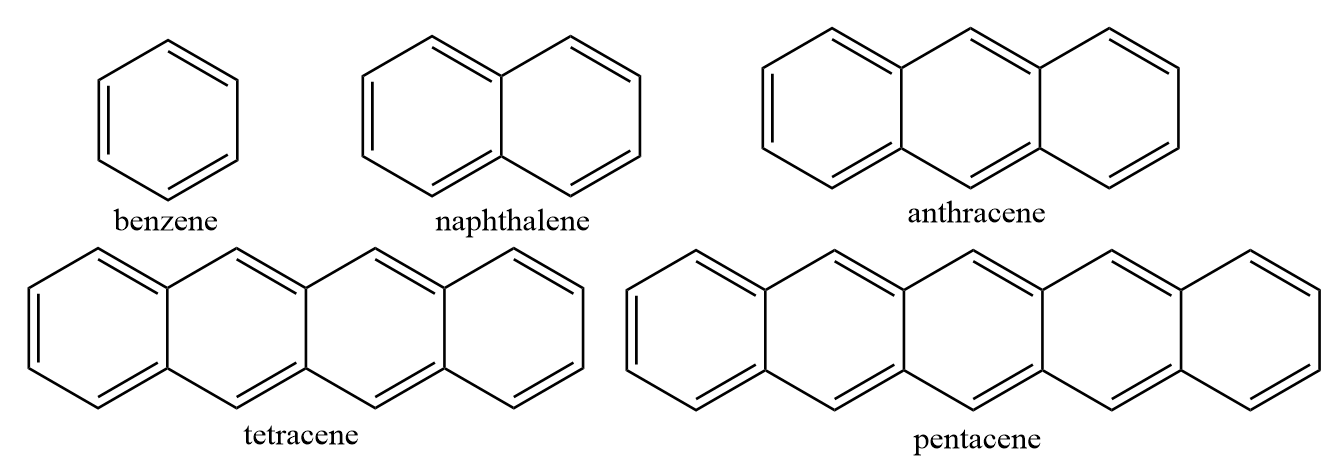} 
    \caption{ Chemical structure of the polyacenes 1-5} 
    \label{figure:polyacenes_chemdraw}
\end{figure}

While the performance for metallocenes is reasonable, the localized orbital approximation is likely to have a greater effect on the accuracy of ph-AFQMC (as well as a better compression efficiency) as more electron pair interactions are approximated. Thus it is important to assess the scaling of the error of the approximation with respect to the system size. To this end, the singlet-triplet (ST) gaps of the  linear polyacenes provide a set of aromatic, quasi one-dimensional systems for which ph-AFQMC has already been shown to be robust.\cite{shee2019singlet} As these systems exhibit significant delocalization throughout their $\pi$ systems, we expect the calculation of this quantity to provide a stringent test of the accuracy and efficiency of LO-AFQMC. Minimal CASSCF trials (6e 6o) were obtained for each system in the cc-pVTZ basis, for both singlet and triplet spin states using the 3 highest and 3 lowest energy occupied and unoccupied RHF orbitals, respectively, following Ref. \citenum{lee2020performance} for benzene. Geometries were taken from Ref. \citenum{shee2019singlet}.

In order to evaluate the error due to the LO approximation, both regular AFQMC and LO-AFQMC with T$_{SVD} = 10^{-4}$ were run for the entire polyacene set. 1656 walkers were run for 300 Ha$^{-1}$, maintaining 99.5$\%$ of the CI weight for each trial wavefunction, and sampling the same auxiliary fields to attempt to remove discrepancies due to rare sampling events. Table \ref{table:polyacene_totalenergies} shows the total energies for each system, as well as the LO error, defined as the difference between the total energy derived from AFQMC and that from LO-AFQMC. In all cases, the LO error remains below 1 kcal/mol, and additionally below that of the statistical error of AFQMC, rendering LO-AFQMC statistically equivalent to AFQMC for these cases. As expected, as the system size increases, the efficacy of localization increases, with pentacene exhibiting a 93.4$\%$ compression. Note that the error does not necessarily cancel between similar species, and additionally does not scale with system size.

\begin{table}[!htb]
    \centering
    \caption{Total energies (Ha) for the singlet (S$_0$) and triplet (T$_1$) states of the series of polyacenes, as calculated by AFQMC and an identical LO-AFQMC calculation with T$_{SVD} = 10^{-4}$.  }
    \begin{tabular}{lr r r r r}
    \hline
System	&	State	&	$\%$ Comp.	&	AFQMC (Ha)	&	LO-AFQMC (Ha)	&	LO Error (kcal/mol)	\\
\hline
benzene	&	S$_0$	&	67	&	-231.9932(5)	&	-231.9928(4)	&	0.2±0.4	\\
	&	T$_1$	&	69	&	-231.8436(6)	&	-231.8439(5)	&	-0.2±0.5	\\
naphthalene	&	S$_0$	&	81	&	-385.4733(6)	&	-385.4739(6)	&	-0.4±0.5	\\
	&	T$_1$	&	81	&	-385.3682(7)	&	-385.3674(6)	&	0.5±0.6	\\
anthracene	&	S$_0$	&	86	&	-538.9477(8)	&	-538.9488(8)	&	-0.7±0.7	\\
	&	T$_1$	&	87	&	-538.8735(10)	&	-538.8741(7)	&	-0.4±0.8	\\
tetracene	&	S$_0$	&	90	&	-692.4214(8)	&	-692.4207(9)	&	0.4±0.8	\\
	&	T$_1$	&	91	&	-692.3684(13)	&	-692.3684(10)	&	0.1±1	\\
pentacene	&	S$_0$	&	94	&	-845.8904(14)	&	-845.8918(11)	&	-0.9±1.1	\\
	&	T$_1$	&	93	&	-845.8555(8)	&	-845.8552(13)	&	0.2±0.9	\\
	
	\hline
    \end{tabular}
    \label{table:polyacene_totalenergies}
\end{table}


The AFQMC and LO-AFQMC singlet triplet gaps for the entire series are presented in Table \ref{table:polyacenes} and compared to previous ph-AFQMC calculations as well as experiment.  In most cases the use of an CASSCF trial, even the minimal 6e 6o ones used here, results in a smaller error compared to experiment, with an MAE of 1.3 $\pm$ 0.8 kcal/mol vs the previously reported ph-AFQMC (using an unrestricted Kohn-Sham (UKS) trial) MAE of 2.8 kcal/mol. LO-AFQMC retains this accuracy with an MAD of 1.8 $\pm$ 0.8 kcal/mol versus experiment, and the LO error ranging from 0.4 to 1.1 kcal/mol, and always within the inherent statistical error of AFQMC.

\begin{table}
\begin{threeparttable} 
\caption{ST gaps in kcal/mol for the polyacenes (cc-pVTZ). The experimental values and ph-AFQMC with a UKS trial were obtained from Ref \citenum{shee2019singlet}; the experimental values are corrected to compare to electronic energies from AFQMC using vibrational free energy corrections from B3LYP as described in Ref \citenum{shee2019singlet}.  } 
\begin{tabular}{l r r r r r}
\hline											
System	&	CAS/AFQMC	&	CAS/LO-AFQMC	&	LO Error	&	UKS/AFQMC\cite{shee2019singlet}	&	Expt.\cite{shee2019singlet}	\\
\hline											
benzene	&	93.9±0.5	&	93.4±0.4	&	-0.4±0.6	&	N/A	&		\\
naphthalene	&	65.9±0.6	&	66.8±0.5	&	0.9±0.8	&	68(1.2)	&	64.4	\\
anthracene	&	46.6±0.8	&	46.9±0.7	&	0.4±1.1	&	46.2(1.2)	&	45.4	\\
tetracene	&	33.2±1	&	32.9±0.8	&	-0.4±1.3	&	34(1.6)	&	31.2	\\
pentacene	&	21.9±1	&	23±1.1	&	1.1±1.4	&	25.2(1.6)	&	21.3	\\
\hline											
MAD vs Expt.	&	1.3±0.8	&	1.8±0.8	&		&	2.8±1.4	&		\\
\hline											
\end{tabular}
\label{table:polyacenes}
\end{threeparttable}
\end{table}

\begin{figure}[H]
    \centering
    \includegraphics[width=8cm]{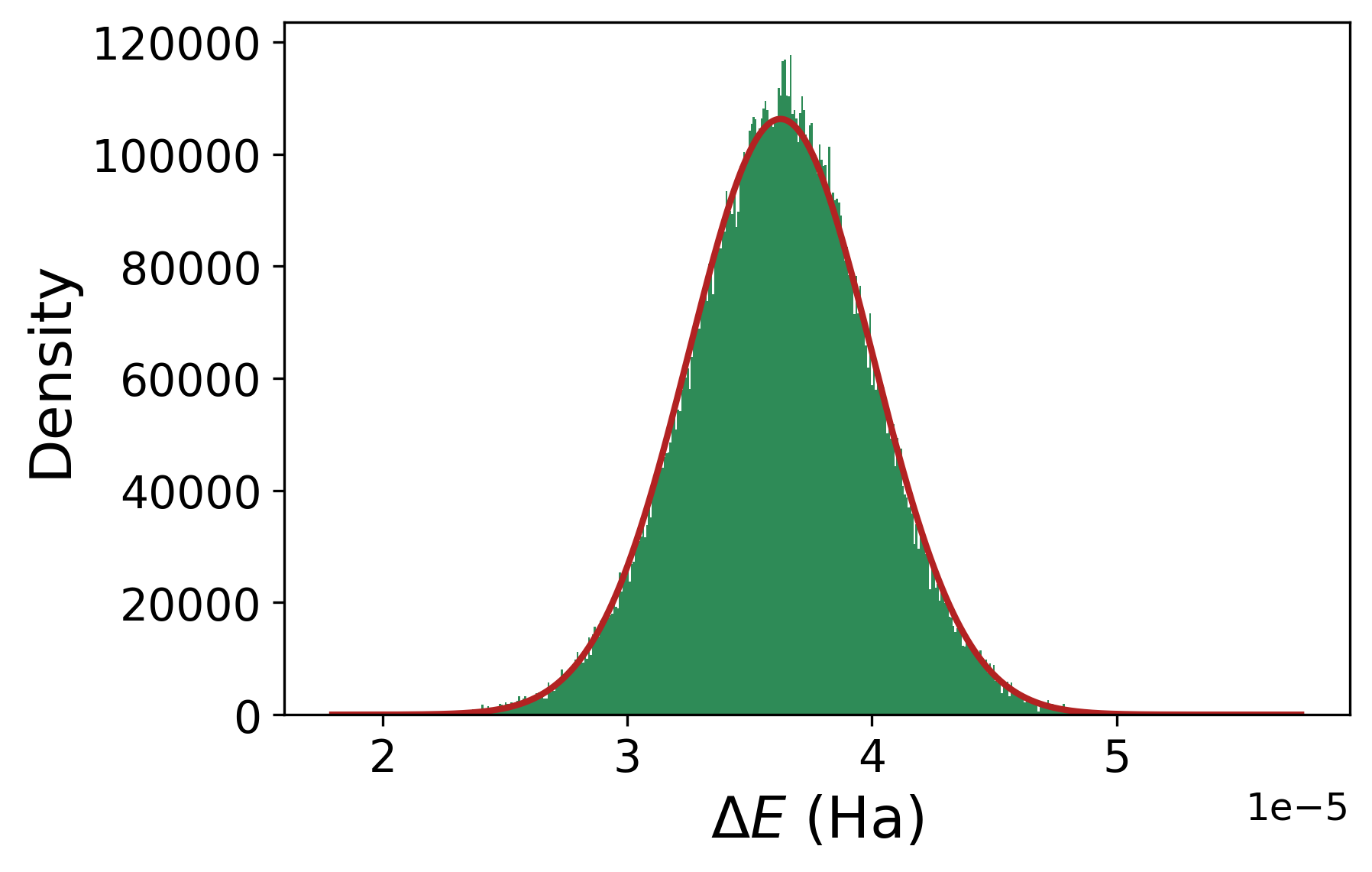} 
    \caption{Distribution of LO error over 100,244 walkers after a single block (0.1 Ha$^{-1}$) of propagation for benzene, fitted to a normal distribution with a mean of 3.6$\times10^{-5}$ and standard deviation of 3.75$\times10^{-6}$. 
    } 
    \label{figure:benzene_hist}
\end{figure}

As we sample walkers along the imaginary time trajectory, there is a small variation in the LO error. To quantify this we take a sample of around 100,000 walkers for the case of benzene, and evaluate the local energy error due to compression (Fig. \ref{figure:benzene_hist}). The error follows a normal distribution at each timestep, which then varies slightly over imaginary time and results in an additional small ($\simeq 0.1$kcal/mol) increase in the statistical error.

\begin{figure}[H]
    \centering
    \includegraphics[width=10cm]{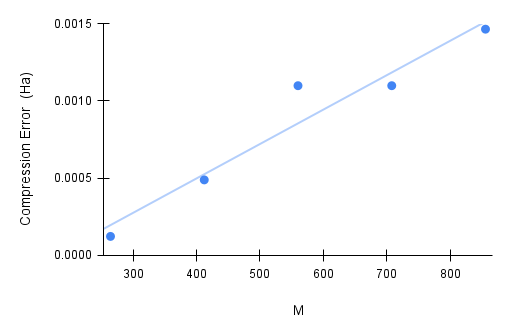} 
    \caption{Scaling of the error of LO-AFQMC vs full ph-AFQMC, estimated at a time slice of 280 Ha$^{-1}$, with respect to the basis size for fixed value of $T_{SVD} = 10^{-4}$. Error bars are not plotted since the error is reported at a single timestep.
    } 
    \label{figure:pac_scaling}
\end{figure}

To investigate whether the insensitivity of the LO error to the size of the molecule was consistent across imaginary time (i.e. independent of the particular set of walkers we ran an energy evaluation using the non-compressed HR-ERI tensor at a single time step (somewhat arbitrarily chosen to be at 280 Ha$^{-1}$), using an identical set of walkers as the LO-AFQMC calculation. The resulting compression errors for the singlet total energies (Fig. \ref{figure:pac_scaling}) do in fact scale near linearly with system size (M), as was originally expected for systems with similar electronic structure. For all systems, the localization error is less than 1 kcal/mol, although for the case of pentacene the error due to compression as estimated at a single time step is actually larger than the statistical error after reblocking over imaginary time. This suggests that the error due to localization, while averaging out to a consistently small value, can vary significantly depending on the stochastic paths taken by the set of walkers. 
Due to this, we believe it may be prudent for applications which depend on the accuracy of the total energy to tighten T$_{SVD}$ as the system size grows, although the scaling of $T_{SVD}$ was not necessary for any of the applications presented in this work. As the efficacy of compression demonstrably increases as the system size increases, we expect the overall savings for percent reduction in memory to remain sizeable, should one tighten T$_{SVD}$ for larger systems. 

\subsection{Platonic Hydrocarbons}

\begin{figure}[H]    
    \begin{center}
        \caption{Scaling of the error of LO-AFQMC vs full ph-AFQMC, estimated at a time step of 2 Ha$^{-1}$ (a) and 250 Ha$^{-1}$ (b), with respect to the basis size (cc-pVTZ). Error bars are not plotted since the error is evaluated at a single timestep. Note that the error for each time step scales linearly, although there is notable variation as the walker distributions change.
    } 
    \label{figure:platonic_scaling}
    \begin{minipage}[b]{0.5\textwidth}
    \centering
    \includegraphics[width=\textwidth]{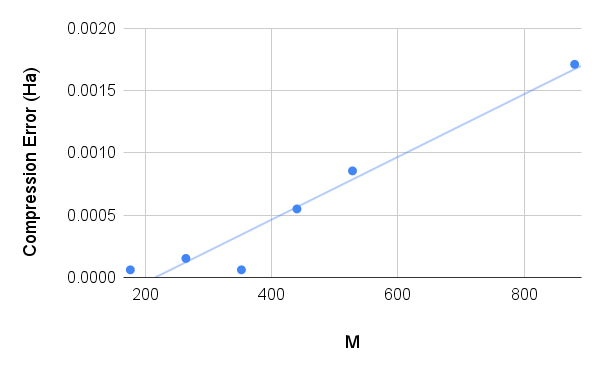}\\
    \subcaption{}
    \end{minipage}%
    \begin{minipage}[b]{0.5\textwidth}
    \centering
    \includegraphics[width=\textwidth]{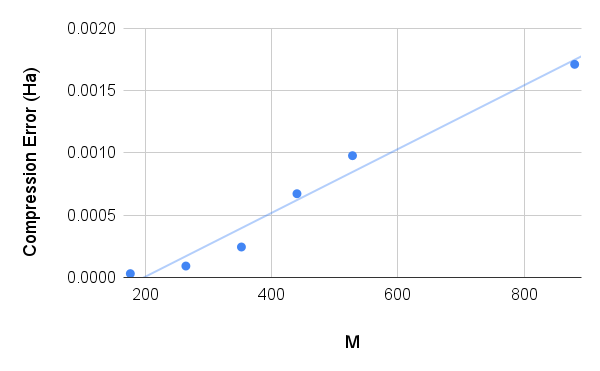}\\
    \subcaption{}
    \end{minipage}
    \end{center}
\end{figure}

While the polyacenes are a good test to observe the scaling of localization error in a quasi one-dimensional setting, it is useful to test three-dimensional systems. Here, we choose a set of hydrocarbon cage molecules in the shape of the platonic solids, for which benchmark CCSD(T) heats of formation and geometries are available.\cite{karton2016heats} Total heats of formation for LO-AFQMC using both cc-pVDZ and cc-pVTZ basis sets are presented in Table \ref{table:platonic_error}, along with LO-AFQMC errors versus full, standard ph-AFQMC calculations and the compression efficiency. Although AFQMC at the cc-pVTZ level reaches the bounds of chemical accuracy versus the cited CCSD(T)/CBS values for the smaller systems, the errors versus the reference CCSD(T) energies increase with the system size. This is not, however, attributed to the LO approximation; the error versus a full ph-AFQMC calculation was in all cases less than the statistical error of ph-AFQMC (LO Error in Table \ref{table:platonic_error}). Two tests (Fig. \ref{figure:platonic_scaling}) using LO-AFQMC and full ph-AFQMC to measure the energy of specific sets of walkers at imaginary times 2 Ha$^{-1}$ and 250 Ha$^{-1}$, representing the beginning and end of an AFQMC simulation respectively, suggest that the linear increase in LO error seen in the polyacenes is indeed reproduced in the platonic hydrocarbons, with some expected variation in error between timesteps. It thus appears that the averaging over imaginary time can again lead to some favorable cancellation of LO error, which is further supported by a small test calculation in the case of \ce{C8H8} in section \ref{sec:SI_error_cancel}. Again, as before, the efficacy of the compression increases with system size, with C$_{20}$H$_{20}$ exhibiting 84$\%$ compression.

\begin{table}
\begin{threeparttable} 
\begin{tabular}{l  r r r r r r }
\hline
	&	CCSD(T)	&	LO-AFQMC	&	LO Error	&	$\%$ Comp.	&	\\
\hline																
	&	CBS	&	cc-pVTZ	&	cc-pVTZ	&	cc-pVTZ	&	\\
\hline																
\ce{C4H4}	&	793.9	&	794.22$\pm$0.34	&	0$\pm$0.48	&	50.4	&	\\
\ce{C6H6}	&	1254.32	&	1255.06$\pm$0.49	&	-0.02$\pm$0.69	&	67.1	&	\\
\ce{C8H8}	&	1706.99	&	1708.58$\pm$0.46	&	-0.15$\pm$0.64	&	72.1	&	\\
\ce{C10H10}	&	2196.46	&	2201.26$\pm$0.5	&	0.76$\pm$0.71	&	82.4	&	\\
\ce{C12H12}	&	2718.29	&	2724.36$\pm$0.56	&	-0.23$\pm$0.8	&	85.9	&	\\
\ce{C20H20}	&	4621.46	&	4633.52$\pm$0.8	&	-0.42$\pm$1.41	&	84.2	&	\\
\hline
\end{tabular}
\end{threeparttable}
\centering
\caption{Total Atomization Energies in kcal/mol for the platonic hydrocarbon cages; CCSD(T)/CBS values were taken from Ref. \citenum{karton2016heats}. LO Error refers to the difference between the LO-AFQMC result and an identical ph-AFQMC calculation run without compression.}\label{table:platonic_error}
\end{table}

\subsection{GPU implementation and timing}
\begin{figure}[H]
    \centering
    \includegraphics[width=10cm]{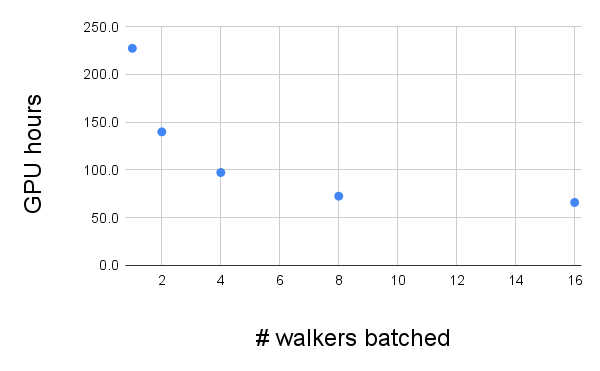} 
    \caption{Effect of performing operations for increasing numbers of walkers in parallel on a single GPU, for Fe(Cp)$_2$ in the cc-pVTZ-DK basis (508 basis functions, 200 determinants). The GPU-hours represent the time necessary to propagate 1656 walkers over 200 Ha$^{-1}$, evaluating the energy 2000 times in total per walker, on the Summit supercomputer at Oak Ridge National Laboratory. 
    } 
    \label{figure:batching}
\end{figure}

LO-AFQMC effectively mediates the memory bottlenecks apparent in the HR-ERI algorithm, in addition to reducing the computational complexity and scaling with respect to system size. However, we have not discussed to what degree the actual wall time can be reduced on a GPU system, as for these system sizes wall time is, more often than not, limited by the ability to perform operations in parallel. Since we are dealing with dense linear algebra, it is possible to perform most operations in parallel for all determinants (aka ``batch'' operations over determinants). As we allow M$_{SVD}$ to vary between each pair $[ij]$, great care must be taken to enable efficient simultaneous batching of operations, especially if one intends to use standard cuBLAS libraries which are designed for calculations with consistent dimension.\cite{cuda} By storing the compressed ERIs contiguously in GPU memory, we were able to batch nearly all operations over the total number of electron pairs and determinants simultaneously, resulting in a significant gain in efficiency. Some details are given in section \ref{sec:SI_GPU}. Additionally, as each walker is undergoing an identical set of operations, we were able to simultaneously evaluate the local energy for all walkers stored on a GPU at once. Such batching is ineffective in the HR-ERI algorithm due to the large memory requirements (and thus inability to store intermediates for each walker), but with the reduced memory requirements, batching LO-AFQMC over walkers significantly improves efficiency, as was previously reported for alternate algorithms in Ref \citenum{malone2020accelerating}. To illustrate this, we ran LO-AFQMC for Fe(Cp)$_2$ in the cc-pVTZ-DK basis, with 200 determinants, 1656 walkers, and propagated to 200 Ha$^{-1}$, while modifying the number of walkers for which operations are performed in parallel on a single GPU. The resulting timings in GPU-hours can be seen in Fig. \ref{figure:batching}; at 16 walkers performed in parallel, the total GPU-hours is reduced by over 3 times versus in series, to 65.9 GPU-hours.

To showcase the speedups possible due to the LO-AFQMC algorithm, as well as the stability of our previous results with respect to use of the LO-AFQMC algorithm, we additionally re-ran the calculations in both cc-pVTZ-DK and cc-pVQZ-DK for the adiabatic ionization energy of Ni(Cp)$_2$ from Ref. \citenum{rudshteyn2021calculation} using $T_{SVD} = 10^{-4}$. For a significantly reduced cost, we were able to reproduce our previous result in the complete basis set (CBS) limit to within statistical error (-141.86$\pm$1.76 kcal/mol versus our previous -143.39$\pm$1.64 kcal/mol), which is additionally within error bars of the experimental result (-143.8$\pm$1.5 kcal/mol). To put this in perspective, our previous ph-AFQMC calculation of the Ni(III)(Cp)$_2$ species in the QZ basis (954 basis functions) required 1615 GPU-hours, whereas LO-AFQMC took 102 GPU hours to propagate the same number of walkers the same distance in imaginary time, reproducing the energy to within statistical error in 6.3$\%$ of the time. Of those 102 GPU hours, approximately 18.9$\%$ were spent in evaluating the energy (both 1- and 2-body terms), slightly more than the approximately 18.7$\%$ that were spent forming the propagation matrix ($B = \sum_{\alpha} x_{\alpha} L_{\alpha}$, scaling as $XM^2$). Although the formation of $B$ is $\simeq$20 times faster than the energy evaluation, it must be performed every step, whereas we only measure the energy once every 20 steps. The compression rates of the neutral/cationic Ni(Cp)$_2$ species were 84/86$\%$ and 86/89$\%$ for the cc-pVTZ-DK and cc-pVQZ-DK bases, respectively, slightly higher than that of Fe(Cp)$_2$.

\section{Conclusions} \label{conclusions}
We have presented a localized orbital compression of the ERIs as a means to reduce the memory footprint and computational scaling of the energy evaluation in AFQMC calculations. Our algorithm has been demonstrated to be efficient and robust for a variety of difficult test systems in manner which is independent of cancellation of error between two electronic states. Although here we restricted our focus to the HR-ERI algorithm of phaseless AFQMC, the approach and approximation errors reported are general. 

As the value of interest is not the controlled error of SVD (namely the L$^2$ norm of the matrix error), but rather the total energy estimated using the compressed integrals, we have taken care to estimate the error due to our localization approximation in multiple ways. We compare to full ph-AFQMC for a single energy evaluation, a weighted average over walkers at a single time step, and full population control AFQMC calculations averaged over imaginary time. In all cases we observe controllable if not entirely monotonic errors for each metric. In addition, we compare our results to experiment and CCSD(T) calculations, highlighting that the accuracy of ph-AFQMC with respect to these external benchmarks remains unaffected due to the LO approximation. 

While the scaling of the algorithms outlined is dependent on the number of electron pairs $N^2$ (i.e.  $\mathcal{O}(N^2(M + N_{det}))$), the formal asymptotic rank of the ERIs is expected to be quadratic $\mathcal{O}(M^2)$, and so we foresee the possibility to improve upon this localization procedure significantly for larger systems. In particular, one might envision screening pairs of electrons, and only including those that contribute meaningfully to the total energy. This would asymptotically reduce the number of pairs to $N$ from $N^2$ and thus reduce the memory scaling of this tensor to quadratic, namely $\mathcal{O}(NM)$, as well as producing a corresponding reduction in computational effort. 

While LO-AFQMC significantly reduces the computational effort required for ph-AFQMC, it is still prohibitively expensive compared to, say, density functional theory (DFT). We thus do not expect LO-AFQMC to directly compete with fast quantum chemistry methods. However, the performance of DFT for transition metal containing molecules has been sub-standard, in no small part due to the lack of trainable reference data for relevant cases. AFQMC can obtain reference values on which to train DFT functionals, and when extended in viable system size via methods like LO-AFQMC, the task of producing a large benchmark dataset with a variety of represented chemistries begins to become feasible. This philosophy is akin to some recent work in the coupled cluster community, where pair-natural orbital (PNO) based localization approaches can reduce the scaling of CCSD(T) to near-linear, with respectable accuracy,\cite{riplinger2013efficient,riplinger2013natural,riplinger2016sparse,datta2016analytic,saitow2017new} and some benchmark datasets on transition metals have been recently produced using DLPNO-CCSD(T).\cite{dohm2018comprehensive,maurer2021assessing} However, for statically correlated systems, or even otherwise, the accuracy of full CCSD(T) can come into question,\cite{hait2019levels} and alternate benchmark-level methods such as AFQMC are necessary to provide complementary or even more accurate results.\cite{shee2019achieving,rudshteyn2020predicting,rudshteyn2021calculation} It is worth noting that a PNO-based ph-AFQMC approach using approximate PNOs from MP2 was initially attempted, but was found to be inaccurate upon projection in imaginary time, presumably due to the approximate nature of the PNO transformation. One advantage of directly compressing the HR-ERI tensor as described here is that it is not dependent on any lower level methods, and so can effectively reproduce full AFQMC values. Even so, the idea to directly solve for a low rank structure for the ERIs is not new to AFQMC, and is shared by tensor hypercontraction and the low-rank compression of the cholesky vectors in Ref. \citenum{motta2019efficient}. The LO-AFQMC algorithm presented here stands out in utility by virtue of it's relative simplicity, high performance, and low prefactor, which depending on the degree of electronic localization within a given molecule can lead to better performance for systems of nearly all sizes. 

Based on the timing results above for the metallocenes test cases, we can estimate the cost of generating AFQMC benchmark data. As an example of a potentially useful benchmark data set, consider the set of experimental redox potentials for octahedral transition metal complexes in solution assembled in Ref. \citenum{hughes2012development}.  The number of heavy atoms in these systems ranges from 7 (e.g. M(H$_2$O)$_6$) to approximately 45; however, most of the molecules have 25 or fewer heavy atoms, and the data set could be restricted to those cases. The cost of running 50 such calculations at the TZ/QZ level would be on the order of 70,000 GPU hours on the Summit supercomputer. A gas phase comparison of DFT and AFQMC results would enable the development of an improved DFT functional for transition metal containing systems; using such a functional, the continuum model needed to compute redox potentials in solution could then be fit to the experimental data assembled in Ref. \citenum{hughes2012development}.

The ability to reproduce full ph-AFQMC for considerably larger molecules than have heretofore been amenable to AFQMC calculations represents a major step towards obtaining benchmark-quality simulations of large transition metal complexes (and other challenging systems) relevant to biology and materials science, such as the water splitting cluster of Photosystem II (PSII). Hundreds to thousands of calculations will be required to obtain relative energies of different metal (Mn) spin and charge states, and different oxygen protonation states, that are needed to precisely specify the intermediates in the Kok cycle, the catalytic process in which the energy of four photons is used to convert water into oxygen and hydrogen gas.\cite{chen2018resolving} A minimal PSII model would contain approximately 50 heavy atoms (based on the smallest possible truncations of the relevant amino acids). The cost of a single calculation at the TZ/QZ level would be on the order of 10,000 GPU hours. While a cost of this magnitude would not be sufficient to evaluate every possible model for the S states of the Kok cycle, 50-100 selected (perhaps on the basis of DFT energetics) individual states could readily be investigated, enabling benchmark relative energies to be determined and compared with various DFT approaches and with experiment.  Application to a system of this size would likely benefit from some of the improvements proposed above, such as screening distant pairs. 

The above applications presume that suitable trial functions can be generated for the systems to be studied, and that a moderate number of determinants (in the 300-1000 range that we have used to date in our AFQMC modeling of metal-containing systems) can produce accurate results.  For the redox potential benchmarks, this assumption is likely to be valid, although there could be surprises for particularly challenging individual cases. For the PSII model, which contains four Mn atoms and one Ca, the requirements for a trial function are far from clear. If an ultralarge number of determinants are needed, it may be necessary to switch to alternate algorithms for better scaling with $N_{det}$ (although, as pointed out above, the LO formulation described in this paper, or some alternate, will still be necessary to make such calculations tractable).

\section{Acknowledgements}

JLW thanks Benjamin Rudshteyn and Shiwei Zhang for many discussions on AFQMC, and Evan Arthur for help with GPU programming. This research used resources of the Oak Ridge Leadership Computing Facility at the Oak Ridge National Laboratory, which is supported by the Office of Science of the U.S. Department of Energy under Contract DE-AC05-00OR22725. This work used the Extreme Science and Engineering Discovery Environment (XSEDE), which is supported by National Science Foundation grant number ACI-1548562. Calculations used the XSEDE resource Expanse at the SDSC through allocation ID COL151. JLW was funded in part by the Columbia Center for Computational Electrochemistry (CCCE). JS acknowledges funding from the National Institute of General Medical Sciences of the National Institutes of Health under award number F32GM142231. 

\section{Associated Content}
The supporting information includes: 
\begin{itemize}
\item Description of the algorithm of Ref. \citenum{mahajan2021taming} and application of the LO approximation
\item Additional Details of the GPU implementation.
\item Discussion of the computational savings due to propagating in the Molecular Orbital Basis.
\item Outline of the SMW algorithm to form the Green's function in the MO basis
\item Data tracking the LO error cancellation for \ce{C8H8} over imaginary time
\end{itemize}
This information is available free of charge via the Internet at http://pubs.acs.org
\newpage
\begin{spacing}{0.85}

\small{
\bibliography{./References.bib} }
\end{spacing}

\newpage

\section{TOC Figure}
\begin{figure}[!htb]
    \includegraphics[height=1.75in]{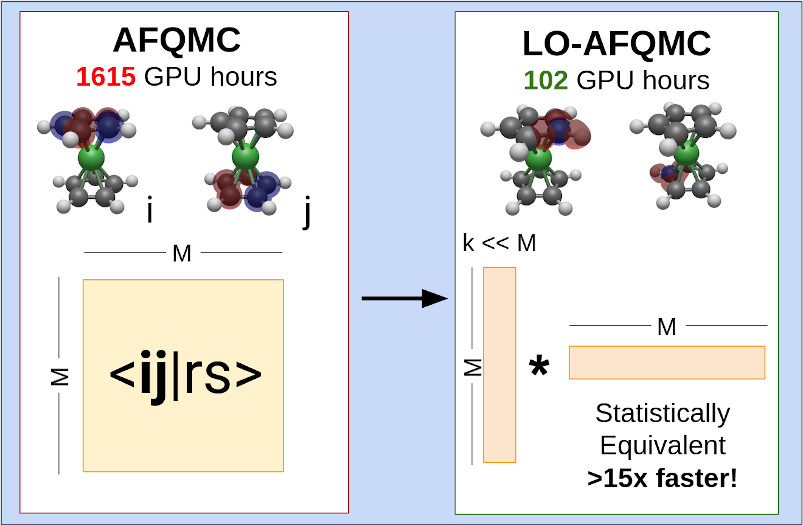} 
    \label{TOC}
\end{figure}

\end{document}